# Spin-Neutral Tunneling Anomalous Hall Effect

Ding-Fu Shao,[1,*,†] Shu-Hui Zhang,[2,*,‡] Rui-Chun Xiao,[3] Zi-An Wang,[1,4] W. J. Lu,[1] Y. P. Sun,[6,1,7] and Evgeny Y. Tsymbal[5,§]

[1] *Key Laboratory of Materials Physics, Institute of Solid State Physics, HFIPS, Chinese Academy of Sciences, Hefei 230031, China*

[2] *College of Mathematics and Physics, Beijing University of Chemical Technology, Beijing, 100029, China*

[3] *Institute of Physical Science and Information Technology, Anhui University, Hefei 230601, China*

[4] *University of Science and Technology of China, Hefei 230026, China*

[5] *Department of Physics and Astronomy & Nebraska Center for Materials and Nanoscience, University of Nebraska, Lincoln, Nebraska 68588-0299, USA*

[6] *High Magnetic Field Laboratory, HFIPS, Chinese Academy of Sciences, Hefei 230031, China*

[7] *Collaborative Innovation Center of Microstructures, Nanjing University, Nanjing 210093, China*

Anomalous Hall effect (AHE) is a fundamental spin-dependent transport property that is widely used in spintronics. It is generally expected that currents carrying net spin polarization are required to drive the AHE. Here we demonstrate that, in contrast to this common expectation, a spin-neutral tunneling AHE (TAHE), i.e. a TAHE driven by spin-neutral currents, can be realized in an antiferromagnetic (AFM) tunnel junction where an AFM electrode with a non-spin-degenerate Fermi surface and a normal metal electrode are separated by a non-magnetic barrier with strong spin-orbit coupling (SOC). The symmetry mismatch between the AFM electrode and the SOC barrier results in an asymmetric spin-dependent momentum filtering of the spin-neutral longitudinal current generating the transverse Hall current in each electrode. We predict a sizable spin-neutral TAHE in an AFM tunnel junction with a $RuO_2$-type AFM electrode and a SnTe-type SOC barrier and show that the Hall currents are reversible by the Néel vector switching. With the Hall angle being comparable to that in conventional AHE bulk materials, the predicted spin-neutral TAHE can be used for the Néel vector detection in antiferromagnetic spintronics.

Spintronics utilizes spin-dependent transport properties of materials for information processing and storage [1]. A typical example is the anomalous Hall effect (AHE), where a longitudinal electric current in a conductor generates a spontaneous transverse Hall current [2-4]. The AHE can be intrinsically driven by the Berry curvature $\mathbf{\Omega}$ [4-6] or induced by extrinsic mechanisms such as spin-dependent impurity scattering in a bulk ferromagnet [4, 7, 8]. In addition, the extrinsic mechanism was predicted to occur in magnetic tunnel junctions due to spin-dependent scattering in a tunneling barrier with strong spin-orbit coupling (SOC) resulting in a tunneling AHE (TAHE) [9-12]. Both the intrinsic and extrinsic AHE are odd under time-reversal symmetry ($\hat{T}$), making this effect useful for the detection of a magnetic order parameter in spintronic devices. However, a ferromagnetic (FM) symmetry group is necessary for a material system to exhibit a finite AHE [4]. This symmetry holds for bulk ferromagnets as well as uncompensated antiferromagnets [13-23], where the presence of a finite net magnetization is not forbidden by $\hat{T}\hat{O}$ symmetry ($\hat{O}$ is a crystal space group symmetry operation). The requirement of FM symmetry implies that the driving and output Hall currents in the AHE conductor are expected to be spin-polarized.

This understanding prohibits the electrical detection of the Néel vector in compensated antiferromagnets by the AHE and thus limits their use in antiferromagnetic (AFM) spintronics [24-26]. This follows from zero net magnetization $\mathbf{M}$ in compensated antiferromagnets, which implies the presence of $\hat{T}\hat{O}$ symmetry ($\hat{T}\hat{O}\mathbf{M} = -\mathbf{M}$) that enforces the Berry curvature $\mathbf{\Omega}$ to be antisymmetric ($\hat{T}\hat{O}\mathbf{\Omega}(\mathbf{k}) = -\mathbf{\Omega}(\hat{T}\hat{O}\mathbf{k})$) in the wave vector $\mathbf{k}$-space and hence the anomalous Hall conductivity (AHC) to vanish. So far, the electrical detection of the Néel vector in AFM devices has been performed using planar Hall-like measurements [27-32]. This approach suffers however from weak signals easily influenced by perturbations [33]. Recently, a non-linear AHE [34,35] was proposed to be used for the Néel vector detection in compensated antiferromagnets [36-38]. While this effect does not require the FM symmetry, and thus both the driving and output currents are spin-neutral, the associated non-linear AHC is typically very small [39]. Thus, it would be interesting from the fundamental point of view and desirable for AFM spintronic applications to find conditions for the emergence of the *spin-neutral AHE*, where the AHC is strong and driven by the spin-neutral currents.

Here, we propose to exploit compensated antiferromagnets with non-spin-degenerate band structures in combination with non-centrosymmetric non-magnetic insulators to realize the spin-neutral AHE. In antiferromagnets with broken $\hat{P}\hat{T}$ ($\hat{P}$ is space inversion) and $\hat{T}\hat{t}$ ($\hat{t}$ is half a unit cell translation) symmetries, the band structure exhibits momentum-dependent spin splitting even in the absence of spin-orbit coupling (SOC) [40-47]. The resulting anisotropic non-spin-degenerate Fermi surface in such antiferromagnets leads to non-trivial spin-dependent transport properties, such as tunneling magnetoresistance in AFM tunnel junctions, driven by the spin-neutral currents [48-50]. On the other hand, non-centrosymmetric materials exhibit intricate spin textures in $\mathbf{k}$-space induced by SOC, which can result in interesting spintronic phenomena [51, 52]. For example, using such a SOC insulator as a barrier in a tunnel junction leads to momentum-dependent spin filtering of a spin-neutral current producing a tunneling spin Hall effect (TSHE) [9-11]. Thus, new spin-controlled transport properties are expected when combing an AFM electrode with a non-spin-degenerate Fermi surface and a non-magnetic barrier layer with strong SOC in a tunnel junction, despite each component in this junction being "spin-neutral" and prohibiting AHE.

In this Letter, based on symmetry analysis and quantum transport calculations, we demonstrate that a spin-neutral TAHE emerges in a tunnel junction consisting of an AFM electrode with a spin-split Fermi surface, a non-magnetic barrier insulator with Ising-type SOC, and a normal metal electrode. In such a junction, a symmetry mismatch between the AFM electrode and the barrier insulator results in an asymmetric spin-dependent momentum filtering of the spin-neutral longitudinal current and generates tunneling AHC (TAHC) and tunneling spin Hall conductance (TSHC) along two orthogonal transverse directions, indicating that the output TAHE current is also spin-neutral. The TAHC changes sign with reversal of the Néel vector in the AFM electrode, and the predicted Hall angle can be comparable to that for a conventional AHE in a magnetic conductor with FM symmetry.

We consider a tunnel junction which consists of magnetic left and normal-metal right electrodes separated by a thin non-magnetic tunneling barrier. For simplicity, the left electrode is assumed to be a collinear magnet with out-of-plane magnetic anisotropy and exchange-split bands. The barrier is assumed to be a non-centrosymmetric two-dimensional (2D) insulator with mirror $\widehat{M}_z$ symmetry, which enforces the Ising-type persistent spin texture at the conduction band minimum with a pure $z$-component [53-55]. In this case, spin is a good quantum number with its axis parallel to the transport $z$-direction. The longitudinal conductance of this tunnel junction is given by [56]

$$G_{zz} = \frac{e^2}{(2\pi)^3 \hbar} \int T_{\mathbf{k}_\parallel} d\mathbf{k}_\parallel, \quad (1)$$

where $T_{\mathbf{k}_\parallel} = \sum_\sigma T_{\mathbf{k}_\parallel}^\sigma$ is the transmission at the transverse wave vector $\mathbf{k}_\parallel = (k_x, k_y)$, and $\sigma$ denotes the spin component ↑ or ↓. The TAHC and TSHC in electrode $i$ ($i = L$ and $i = R$ for left and right electrodes, respectively) of this junction are

$$G_{\eta z, i} = \frac{e^2}{(2\pi)^3 \hbar} \int \left( \frac{v_{\eta,i}^\uparrow}{|v_{z,i}^\uparrow|} T_{\mathbf{k}_\parallel}^\uparrow + \frac{v_{\eta,i}^\downarrow}{|v_{z,i}^\downarrow|} T_{\mathbf{k}_\parallel}^\downarrow \right) d\mathbf{k}_\parallel,$$

$$G_{\eta z, i}^z = \frac{e^2}{(2\pi)^3 \hbar} \int \left( \frac{v_{\eta,i}^\uparrow}{|v_{z,i}^\uparrow|} T_{\mathbf{k}_\parallel}^\uparrow - \frac{v_{\eta,i}^\downarrow}{|v_{z,i}^\downarrow|} T_{\mathbf{k}_\parallel}^\downarrow \right) d\mathbf{k}_\parallel, \quad (2)$$

where $v_{r,i}^\sigma = \frac{\partial E_i^\sigma}{\hbar \partial k_r}$ is the band velocity along the $r$ direction, and $\eta = x, y$ is the transverse direction. The spin polarization of TAHC can be then defined as $p_{\eta z, i} = \frac{G_{\eta z,i}^z}{G_{\eta z,i}}$.

For the right electrode described by a free-electron model where $v_{r,R}^\sigma = v_{r,R} = \frac{\hbar}{m} k_r$, Eq. (2) can be written as [11,12]

$$G_{\eta z, R} = \frac{e^2}{(2\pi)^3 \hbar} \int \frac{k_{\eta,R}}{|q_{z,R}|} T_{\mathbf{k}_\parallel} d\mathbf{k}_\parallel,$$

$$G_{\eta z, R}^z = \frac{e^2}{(2\pi)^3 \hbar} \int \frac{k_{\eta,R}}{|q_{z,R}|} p_{\mathbf{k}_\parallel} T_{\mathbf{k}_\parallel} d\mathbf{k}_\parallel. \quad (3)$$

Here $q_{z,R}$ is the $z$-component of the Fermi wave vector of the right electrode, and $p_{\mathbf{k}_\parallel} = \frac{T_{\mathbf{k}_\parallel}^\uparrow - T_{\mathbf{k}_\parallel}^\downarrow}{T_{\mathbf{k}_\parallel}}$ is the $\mathbf{k}_\parallel$-dependent transport spin polarization. Since $\frac{k_{\eta,R}}{|q_{z,R}|}$ is an odd function of $\mathbf{k}_\parallel$, $G_{\eta z, R}$ and $G_{\eta z, R}^z$ are non-zero only when $T_{\mathbf{k}_\parallel}$ and $p_{\mathbf{k}_\parallel} T_{\mathbf{k}_\parallel}$ are uneven, i.e. their distributions in $\mathbf{k}_\parallel$-space are asymmetric. In the ballistic transport regime with no spin-flip scattering, spin $\sigma$ and wave vector $\mathbf{k}_\parallel$ are conserved, and $T_{\mathbf{k}_\parallel}^\sigma$ is determined by the matching between the spin-dependent barrier height $U_{\mathbf{k}_\parallel}^\sigma$ and the number of conduction channels $N_{\mathbf{k}_\parallel,i}^\sigma$ in each electrode. The latter has spin polarization $p_{\mathbf{k}_\parallel,i} = \frac{N_{\mathbf{k}_\parallel,i}^\uparrow - N_{\mathbf{k}_\parallel,i}^\downarrow}{N_{\mathbf{k}_\parallel,i}^\uparrow + N_{\mathbf{k}_\parallel,i}^\downarrow}$. For the former, we define a $\mathbf{k}_\parallel$-dependent spin filtering factor $f_{\mathbf{k}_\parallel} = -\frac{U_{\mathbf{k}_\parallel}^\uparrow - U_{\mathbf{k}_\parallel}^\downarrow}{U_{\mathbf{k}_\parallel}^\uparrow + U_{\mathbf{k}_\parallel}^\downarrow}$, where a positive (negative) $f_{\mathbf{k}_\parallel}$ implies a lower (higher) barrier height for up-spin electrons. Since $p_{\mathbf{k}_\parallel,R} = 0$ for the normal-metal electrode, the distribution of $T_{\mathbf{k}_\parallel}$ is determined by the matching of $p_{\mathbf{k}_\parallel,L}$ and $f_{\mathbf{k}_\parallel}$.

Figure 1(a) schematically shows $f_{\mathbf{k}_\parallel}$ for the insulator with the Ising-type SOC texture. Time reversal symmetry $\widehat{T}$ enforces antisymmetry of $f_{\mathbf{k}_\parallel}$, i.e. $\widehat{T} f_{\mathbf{k}_\parallel} = -f_{-\mathbf{k}_\parallel}$. Usually, there exist additional crystal symmetry, e.g., mirror $\widehat{M}_x$ symmetry for the point group $mm2$ as in 2D SnTe [57,58], which requires $\widehat{M}_x f_{k_x} = -f_{-k_x}$, as depicted in Fig. 1(a). If the left electrode is a FM metal, $\widehat{T}$ symmetry is broken, while $\widehat{T}\widehat{U}_s$ symmetry ($\widehat{U}_s$ is spin rotation) holds enforcing $\widehat{T}\widehat{U}_s p_{\mathbf{k}_\parallel,L} = p_{-\mathbf{k}_\parallel,L}$. Normally, in FM metals, $p_{\mathbf{k}_\parallel,L}$ has the same sign in all quadrants of the 2D Brillouin zone (e.g., due to four-fold rotational symmetry), as shown in Fig. 1(b), resulting in a finite global spin polarization $p_L = \frac{\sum_{\mathbf{k}_\parallel} \left( N_{\mathbf{k}_\parallel,L}^\uparrow - N_{\mathbf{k}_\parallel,L}^\downarrow \right)}{\sum_{\mathbf{k}_\parallel} \left( N_{\mathbf{k}_\parallel,L}^\uparrow + N_{\mathbf{k}_\parallel,L}^\downarrow \right)}$.

In a tunnel junction composed of a SOC barrier with $f_{\mathbf{k}_\parallel}$ as in Fig. 1(a) and a FM electrode with $p_{\mathbf{k}_\parallel,L}$ as in Fig. 1(b), $T_{\mathbf{k}_\parallel}$ is suppressed for $k_x < 0$. This is due to the mismatch of $f_{\mathbf{k}_\parallel}$ and $p_{\mathbf{k}_\parallel,L}$ filtering out transverse transmission contributions with negative $k_x$ and thus generating Hall and spin Hall currents along the $x$ direction (Fig. 1(d)). In this case, both $p_L$ and $p_{xz,R}$ are nonvanishing, i.e. both the input and output TAHE currents are spin-polarized. The spin-polarized nature of the TAHE in this case is consistent with previous understanding of the extrinsic AHE due to skew scattering [4-12].

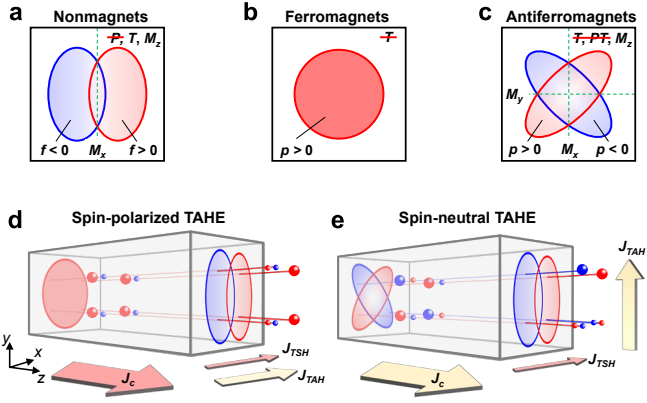
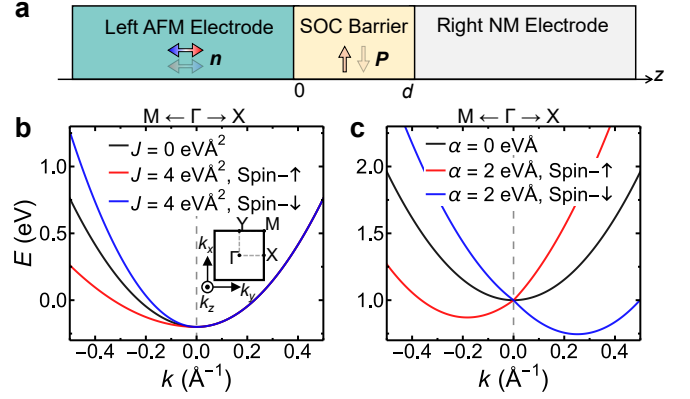

**Fig. 1** (a-c)Schematics of the $k_\parallel$-dependent spin filtering factor $f_{k_\parallel}$ for a SnTe-type SOC barrier with the Ising-type persistent spin texture (a), spin polarization $p_{k_\parallel,L}$ for a FM metal (b), and for a RuO$_2$-type AFM metal with spin-split Fermi surface (c). (d) Schematics of the spin-polarized TAHE in a FM tunnel junction, where a spin-polarized current from the FM electrode with $p_{k_\parallel,L}$ as in (b) tunnels across the SOC barrier with $f_{k_\parallel}$ as in (a), generating the transverse charge and spin currents in the same direction. (e) Schematics of the spin-neutral TAHE in an AFM tunnel junction, where a spin-neutral current from the AFM electrode with $p_{k_\parallel,L}$ as in (c) tunnels across the SOC barrier with $p_{k_\parallel,B}$ as in (a), generating the transverse charge and spin currents in orthogonal directions.

**Fig. 2** (a) Schematic of the AFM tunnel junction considered in this work. The out-of-plane Néel vector $n$ of the AFM electrode and the in-plane ferroelectric polarization $P$ in the SOC barrier are indicated by arrows. (b,c) Bulk band structures of the left AFM electrode (b) and the SOC barrier (c). The inset in (b) depicts the top view of the Brillouin zone.

Next, we consider an AFM metal with a spin-split Fermi surface as the left electrode. Figure 1(c) shows spin polarization $p_{k_\parallel,L}$ for such an antiferromagnet assuming the presence of mirror symmetries $\hat{M}_x$, $\hat{M}_y$, and $\hat{M}_z$ in its magnetic point group (4'/mm'm). This is the case for the recently discovered rutile-type antiferromagnets with an out-of-plane Néel vector, such as RuO$_2$ [59,60], where the symmetry transformation $\hat{M}_z E^\sigma_{k_\parallel,k_z} = E^\sigma_{k_\parallel,-k_z}$ conserves spin $\sigma = \uparrow, \downarrow$ of the conduction modes at $k_\parallel$ and there is no symmetry operation enforcing $p_{k_\parallel,L} = 0$ [48]. $\hat{T}\hat{U}_s$ symmetry is preserved in the absence of SOC, resulting in $\hat{T}\hat{U}_s p_{k_\parallel,L} = p_{-k_\parallel,L}$. Mirror symmetries $\hat{M}_x$ and $\hat{M}_y$ lead to $\hat{M}_x p_{k_x,k_y,L} = -p_{-k_x,k_y,L}$ and $\hat{M}_y p_{k_x,k_y,L} = -p_{k_x,-k_y,L}$, which imply a zero net spin polarization $p_L$.

Combining a RuO$_2$-type AFM electrode and a SnTe-type SOC barrier in a tunnel junction leads to a matching (mismatching) of $p_{k_\parallel,L}$ and $f_{k_\parallel}$ for $k_y > 0$ ($k_y < 0$) (Fig. 1(e)). As a result, $T_{k_\parallel}$ is enhanced (reduced) for positive (negative) $k_y$ making it asymmetric in the $k_y$-direction and thus TAHC $G_{yz,R}$ finite. Therefore, a spin-neutral current from the AFM electrode can drive a TAHE due to asymmetric spin-dependent momentum filtering. On the other hand, due to $\hat{M}_x$ symmetry of both the AFM electrode and the SOC barrier, $T_{k_\parallel}$ is symmetric with respect to $\hat{M}_x$ resulting in zero $G_{xz,R}$. In contrast, $p_{k_\parallel} T_{k_\parallel}$ is antisymmetric with respect to $\hat{M}_x$ but symmetric with respect to $\hat{M}_y$, leading to finite $G^z_{xz,R}$ and zero $G^z_{yz,R}$. The resulting $p_{yz,R} = \frac{G^z_{yz,R}}{G_{yz,R}} = 0$ indicates that not only the driving current but also the output Hall current is spin-neutral. This is contrary to either intrinsic or extrinsic AHE, both being controlled by spin-polarized currents, considered previously.

In order to quantify the spin-neutral TAHE, we consider an AFM tunnel junction that consists of a semi-infinite RuO$_2$-type left AFM electrode ($z < 0$) and a right nonmagnetic electrode ($z > d$) separated by a SnTe-type SOC barrier of thickness $d$ (Fig. 2(a)). The corresponding Hamiltonians $H_L$, $H_B$, and $H_R$ for the left electrode, the barrier layer, and the right electrode are

$$H_L = -\frac{\hbar^2}{2m}\nabla^2 + V_L + J\sigma_z k_x k_y,$$
$$H_B = -\frac{\hbar^2}{2m}\nabla^2 + V_B + \alpha \sigma_z k_x,$$
$$H_R = -\frac{\hbar^2}{2m}\nabla^2 + V_R. \quad (4)$$

Here $m$ is the electron mass, $\sigma_z$ is the Pauli matrix, $V_L$, $V_B$, and $V_R$ are potentials in the left electrode, barrier layer, and right electrode, respectively, $J$ is the exchange splitting parameter, and $\alpha$ is the SOC parameter which is ~1.2 eVÅ for SnTe [57].

We choose $V_L = -0.2$ eV characterizing the band structure of RuO$_2$ near the Fermi level $E_F$ [23,48], $V_B = 1$ eV representing a typical band gap of SnTe-type chalcogenides [57], and $V_R = -1$ eV to describe a normal metal. The band structure of the AFM electrode (Fig. 2(b)) has a large spin splitting except the Γ-X and Γ-Y directions in the Brillouin zone due to the $\hat{M}_x$ and $\hat{M}_y$ symmetries of magnetic point group 4'/mm'm. On the other hand, the electronic bands along the Γ-X direction are non-spin-degenerate for the SOC barrier (Fig. 2(c)), due to the absent

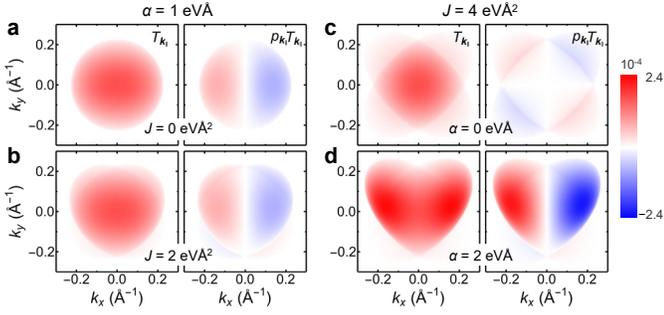

**Fig. 3** (a,b) Calculated $T_{\boldsymbol{k}_\parallel}$ and $p_{\boldsymbol{k}_\parallel}T_{\boldsymbol{k}_\parallel}$ as a function of $\boldsymbol{k}_\parallel$ for the AFM tunnel junction for $\alpha = 1$ eVÅ with $J = 0$ (a) and $J = 2$ eVÅ$^2$ (b). (c,d) The same as in (a) but for $J = 4$ eVÅ$^2$ with $\alpha = 0$ (c) and $\alpha = 2$ eVÅ (d).

$\widehat{M}_y$ symmetry in SnTe-type chalcogenides. Such band structures result in $p_{\boldsymbol{k}_\parallel,L}$ and $f_{\boldsymbol{k}_\parallel}$ schematically shown in Figures 1(a,c).

We calculate TAHC and TSHC for this AFM tunnel junction as described in Supplemental Material [61]. In the calculation, we assume $d = 1$ nm, which is about the thickness of one or two atomic layers for 2D materials. Figure 3 shows the calculated $T_{\boldsymbol{k}_\parallel}$ and $p_{\boldsymbol{k}_\parallel}T_{\boldsymbol{k}_\parallel}$ at $E_F = 0$ eV. For a non-zero SOC parameter, as $\alpha = 1$ eVÅ in Fig. 3(a), we find that in the absence of exchange splitting ($J = 0$), $T_{\boldsymbol{k}_\parallel}$ is isotropic. Therefore, $G_{\eta z,R}$ is zero in the right electrode (Fig. 4(b)). On the other hand, $p_{\boldsymbol{k}_\parallel}T_{\boldsymbol{k}_\parallel}$ is antisymmetric with respect to $\widehat{M}_x$ (Fig. 3(a)) resulting in sizable $G^z_{xz,R}$ but vanishing $G^z_{yz,R}$ (Fig. 4(c)). When $J$ is finite, $N^\sigma_{\boldsymbol{k}_\parallel,L}$ becomes anisotropic, and areas with nonzero $p_{\boldsymbol{k}_\parallel,L}$ appear at $\boldsymbol{k}_\parallel$ away from the $\widehat{M}_x$ and $\widehat{M}_y$ invariant lines, as schematically shown in Fig. 1(c). In this case, $T_{\boldsymbol{k}_\parallel}$ is suppressed in such areas with $k_y < 0$, due to the mismatch of $p_{\boldsymbol{k}_\parallel,L}$ and $p_{\boldsymbol{k}_\parallel,B}$. The asymmetry of $T_{\boldsymbol{k}_\parallel}$ along the $y$ direction (Fig. 3(b)) leads to a finite $G_{yz,R}$ (Fig. 4(b)). $G_{xz,R}$ remains zero since $T_{\boldsymbol{k}_\parallel}$ is still symmetric with respect to $\widehat{M}_x$.

Hall and spin Hall angles, $\frac{G_{yz,R}}{G_{zz}}$ and $\frac{G^z_{xz,R}}{G_{zz}}$ respectively, can serve as a figure of merit of the predicted TAHE and TSHE. We find that $\frac{G_{yz,R}}{G_{zz}} > 1\%$ for $J > 1$ eVÅ$^2$, which can be enhanced to 4% for $J = 4$ eVÅ$^2$. Such a sizable Hall angle is comparable to that usually found in bulk magnets [4]. It is notable that $G_{yz,R}$ is an odd function of $J$ (Fig. 4(b)), indicating that TAHE is reversed with switching of the Néel vector. On the other hand, $G^z_{xz,R}$ is an even function of $J$, and thus the spin Hall angle $\frac{G^z_{xz,R}}{G_{zz}} \sim 10\%$ is almost independent of $J$ (Fig. 4(c)), demonstrating that the TSHE is majorly controlled by the SOC barrier rather than the AFM electrode.

Figures 3(c,d) and 4(d,e,f) show the effect of SOC (for $J = 4$ eVÅ$^2$). When $\alpha = 0$, both $T_{\boldsymbol{k}_\parallel}$ and $p_{\boldsymbol{k}_\parallel}T_{\boldsymbol{k}_\parallel}$ are symmetric in $\boldsymbol{k}_\parallel$ resulting in zero $G_{\eta z,R}$ and $G^z_{\eta z,R}$ (Figs. 3(c) and 4(e,f)). For a

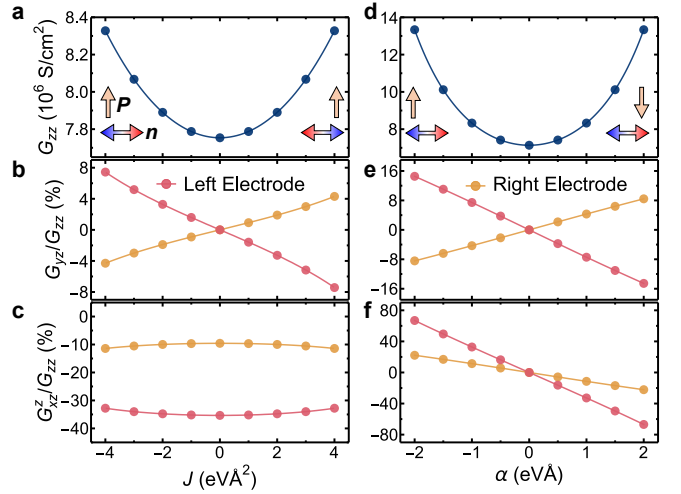

**Fig. 4** (a,b,c) The calculated $G_{zz}$ (a), $G_{yz}/G_{zz}$ (b) and $G^z_{xz}/G_{zz}$ (c) in the left and right electrodes for the AFM tunnel junction with $\alpha = 1$ eVÅ and variable $J$. (d,e,f) The calculated $G_{zz}$ (d), $G_{yz}/G_{zz}$ (e) and $G^z_{xz}/G_{zz}$ (f) in the left and right electrodes for the AFM tunnel junction with $J = 4$ eVÅ$^2$ and variable $\alpha$. Arrows in (a) denote schematically the orientations of the Néel vector $\boldsymbol{n}$ and ferroelectric polarization $\boldsymbol{P}$ in the AFM tunnel junction shown in Fig. 2(a).

finite $\alpha$, spin-dependent momentum filtering makes $T_{\boldsymbol{k}_\parallel}$ asymmetric along the $y$ direction resulting in a finite $G_{yz,R}$ (Figs. 3(d) and 4(e)). We find that the $G_{yz,R}$ and $G^z_{xz,R}$ are both enhanced linearly with the increasing $\alpha$ (Fig. 4(b)). For a large $\alpha$ of 2 eVÅ, we obtain spectacular heart-shape patterns for $T_{\boldsymbol{k}_\parallel}$ and $p_{\boldsymbol{k}_\parallel}T_{\boldsymbol{k}_\parallel}$ (Fig. 3(d)) and sizable Hall angle $\frac{G_{yz,R}}{G_{zz}} \sim 8\%$ and spin Hall angle $\frac{G^z_{xz,R}}{G_{zz}} \sim 16\%$ (Figs. 4(e,f)).

We note that SnTe-type insulators exhibit an in-plane ferroelectric polarization [62]. Similar to other ferroelectrics [63-68], the polarization is coupled to the SOC parameter $\alpha$ which determines their spin texture in **k**-space [52]. Polarization switching reverses the spin texture and hence $f_{\boldsymbol{k}_\parallel}$ [57, 58]. As a result, both $G_{yz,R}$ and $G^z_{xz,R}$ are odd functions of $\alpha$ (Figs. 4(e,f)), indicating that the TAHE and TSHE are reversible by ferroelectric switching. Therefore, ferroelectric polarization of the barrier layer can serve as an additional control parameter of the tunneling anomalous Hall and spin Hall effects.

Hall and spin Hall currents also emerge in the left AFM electrode. While, similar to the right electrode, $G_{xz,L}$ and $G^z_{yz,L}$ vanish by symmetry, $G_{yz,L}$ and $G^z_{xz,L}$ are larger in magnitude than their counterparts $G_{yz,R}$ and $G^z_{xz,R}$ (Fig. 4), due to the smaller longitudinal Fermi velocities in the denominators of Eq. (2) for the left electrode. It is notable that $G_{yz,L}$ and $G_{yz,R}$ have opposite sign (Fig. 4(b,e)), which is due to additional contributions from transverse velocities induced by the exchange splitting (Eq. (4)). The detailed discussion is given in Supplemental Material [61].

The predicted spin-neutral TAHE is feasible in experiment. Several AFM metals with non-spin-degenerate Fermi surfaces have been found recently [40-45,59,60]. Especially, high-quality epitaxial films of AFM $RuO_2$ grown experimentally [69-71] are suitable to serve as an AFM electrode in the proposed tunnel junction. The Ising-type spin textures widely exist in non-centrosymmetric 2D materials with $\hat{M}_z$ symmetry, such as monochalcogenide MX and H-phase dichalcogenide $MX_2$ monolayers (M = metals and X = S, Se, or Te) [53-55,57,58]. It would be straightforward to attach such an exfoliated monolayer SOC barrier on $RuO_2$ for the experimental observation of the spin-neutral TAHE. Twisting between the SOC monolayer and $RuO_2$ could be used to break $\hat{M}_x$ and $\hat{M}_y$ symmetries in both the electrodes and the barrier. This would allow generation of the spin-polarized output Hall current from the input spin-neutral longitudinal current, unexpected by the previous understanding of either intrinsic or extrinsic AHE where both the input and output TAHE currents must be spin-polarized.

It is also possible to use other types of AFM electrodes and SOC barriers to realize the spin-neutral TAHE. Several SOC compounds with in-plane persistent spin textures have been predicted recently [72,73] and could be used to match the non-spin-degenerate conduction channels of the AFM electrode with in-plane anisotropy. In addition, the barrier layer can also have conventional Rashba- or Dresselhaus-type spin textures [74,75]. In this case, although spin is not a good quantum number, the asymmetric $T_{\boldsymbol{k}_\parallel}$ can be still generated for a finite TAHE [12], due to the mismatch of $\hat{T}$ and $\hat{T}\hat{U}_s$ symmetries of the electrodes and the barrier layer in the AFM tunnel junction.

The proposed spin-neutral TAHE offers promising opportunities for AFM spintronics based on single-layer antiferromagnets. The odd nature of TAHC with respect to $J$ allows the detection of the Néel vector reversal by measuring the transverse Hall voltage on either side of the proposed AFM tunnel junction. The large Hall angle indicates a strong signal robust to perturbations. Thus, the proposed approach has significant advantages compared to the planar Hall-like measurements [27-33] widely used for the Néel vector detection in antiferromagnets. We hope therefore that our predictions will stimulate experimental explorations of the TAHE driven by spin-neutral currents.

**Acknowledgments.** This work was supported by the National Science Foundation of China (NSFC Grants No. 12174019, U2032215), E.Y.T. acknowledges support from the EPSCoR RII Track-1 program (NSF Award OIA-2044049). The calculations were performed at Hefei Advanced Computing Center. The figures were created using the SciDraw scientific figure preparation system [76].

\* These authors contributed equally to this work.

† dfshao@issp.ac.cn

‡ shuhuizhang@mail.buct.edu.cn

§ tsymbal@unl.edu